\documentclass[preprint,12pt]{elsarticle}
\usepackage{amssymb}
\usepackage{amsmath}
\usepackage{booktabs}
\usepackage{xspace}

\def\LOfull{${\rm LO}$\xspace}
\def\Bornone{${\rm Born}_1$\xspace}
\def\Borntwo{${\rm Born}_2$\xspace}
\def\Bornthree{${\rm Born}_3$\xspace}
\def\Bornfour{${\rm Born}_4$\xspace}
\def\Bornfive{${\rm Born}_5$\xspace}
\def\Bornfull{${\rm Born}$\xspace}
\def\NLOone{${\rm NLO}_1$\xspace}
\def\NLOtwo{${\rm NLO}_2$\xspace}
\def\NLOthree{${\rm NLO}_3$\xspace}
\def\NLOfour{${\rm NLO}_4$\xspace}
\def\NLOfive{${\rm NLO}_5$\xspace}
\def\NLOsix{${\rm NLO}_6$\xspace}
\def\NLOqcd{${\rm NLO}_{\rm QCD}$\xspace}
\def\NLOfull{${\rm NLO}$\xspace}

\def\asLOone{$\mathcal{O}(\alpha_s^4\alpha^2)$\xspace}
\def\asLOtwo{$\mathcal{O}(\alpha_s^3\alpha^3)$\xspace}
\def\asLOthree{$\mathcal{O}(\alpha_s^2\alpha^4)$\xspace}
\def\asLOfour{$\mathcal{O}(\alpha_s^1\alpha^5)$\xspace}
\def\asLOfive{$\mathcal{O}(\alpha_s^0\alpha^6)$\xspace}

\def\asNLOone{$\mathcal{O}(\alpha_s^5\alpha^2)$\xspace}
\def\asNLOtwo{$\mathcal{O}(\alpha_s^4\alpha^3)$\xspace}
\def\asNLOthree{$\mathcal{O}(\alpha_s^3\alpha^4)$\xspace}
\def\asNLOfour{$\mathcal{O}(\alpha_s^2\alpha^5)$\xspace}
\def\asNLOfive{$\mathcal{O}(\alpha_s^1\alpha^6)$\xspace}
\def\asNLOsix{$\mathcal{O}(\alpha_s^0\alpha^7)$\xspace}
\journal{Journal of Subatomic Particles and Cosmology}

\begin{document}

\begin{frontmatter}
\title{Full off-shell $t\bar{t}$ production in the semileptonic channel \\ at the LHC}

\author[a]{Leon Mans}
\ead{leon.mans@rwth-aachen.de}

\affiliation[a]{Institute for Theoretical Particle Physics
and Cosmology, RWTH Aachen University, D-52056 Aachen, Germany}

\begin{abstract}
In this contribution, we summarize our results on the complete NLO predictions for the production of off-shell top quark pairs in the $\ell+j$ channel at the LHC. All NLO QCD and electroweak corrections are consistently included  across the full set of LO contributions and partonic subprocesses, while retaining finite-width effects of the top quarks and the electroweak gauge bosons. Resonant and non-resonant contributions as well as interference effects are included in the complete calculation. We pay special attention to infrared singularities associated with photons and jets appearing simultaneously in the final state. Fiducial integrated and differential cross-section results are presented for $pp$ collisions at $\sqrt{s}=13.6$ TeV.
\end{abstract}


\end{frontmatter}

\section{Introduction}
\label{sec:introduction}

Top-quark pair production in the $\ell+j$ channel remains one of the most important processes for precision studies at the LHC. Owing to its relatively large branching ratio, this channel provides a higher statistical precision compared to the dileptonic channel, while the presence of a single neutrino allows for efficient background suppression through kinematic reconstruction techniques. As a result, it plays a central role in tests of the SM and in searches for indirect signals of physics beyond the SM.

A reliable theoretical description of this process requires the inclusion of full off-shell effects of the top quarks and gauge bosons. Specifically, all double-, single-, and non-resonant contributions should be included together with interferences, and the finite-width effects of the unstable particles. Such effects become particularly relevant in the phase-space regions with large transverse momenta, which are of special interest for new-physics searches. The latter regions will be increasingly accessible with the High-Luminosity upgrade of the LHC. So far, theoretical predictions for this channel have been limited to NLO QCD corrections \cite{Denner:2017kzu}.

In this contribution, we summarize our complete NLO calculation for off-shell top-quark pair production in the $\ell+j$ channel, including the full set of NLO QCD and electroweak corrections to all contributing Born-level amplitudes \cite{Mans:2025czh}. To ensure IR and collinear safety, we implement quark-to-photon fragmentation functions \cite{Denner:2014bna,Stremmer:2024zhd} and photon-to-jet conversion functions \cite{Denner:2019zfp}. We report the main results that have been presented in Ref.~\cite{Mans:2025czh}, describe the methods employed, and provide some examples of fiducial integrated and  differential cross-section results at various perturbative orders, while highlighting the impact of subleading effects.

\section{Process definition and computational framework}
\begin{figure}[t!]
    \centering
    \includegraphics[width=\linewidth]{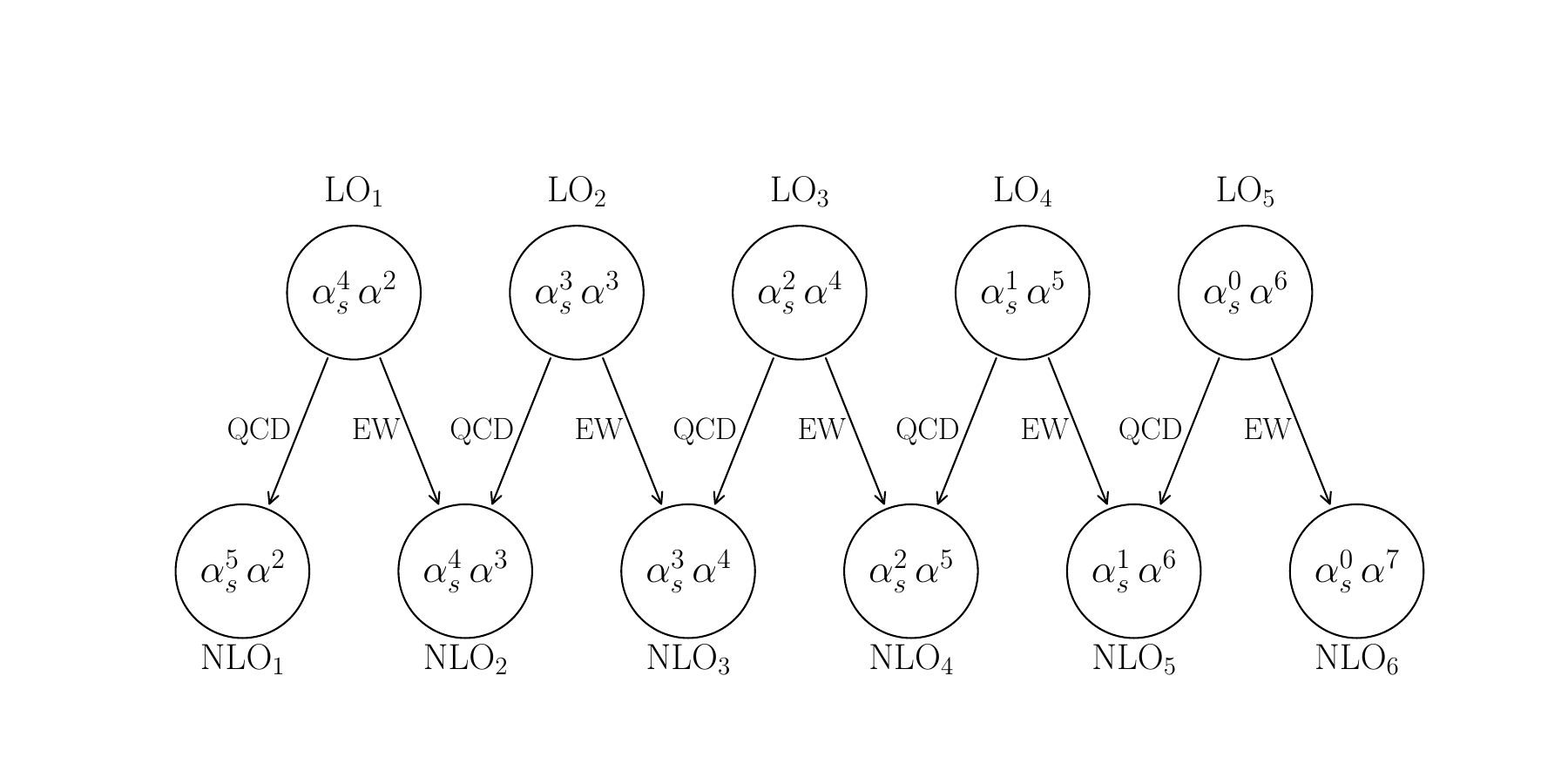}
    \caption{\it Diagrammatic overview of the LO contributions and their corresponding NLO corrections for
$pp \to \ell^- \bar{\nu}_\ell\, j_b j_b\, jj + X$.}

    \label{fig:orders}
\end{figure}

All relevant perturbative orders are visualized in Figure \ref{fig:orders}. At leading order, the complete prediction is organized into five perturbative orders, ranging from the QCD-dominated $\mathcal{O}(\alpha_s^4\alpha^2)$ contribution down to the purely EW-induced $\mathcal{O}(\alpha^6)$ contribution, including photon-initiated subprocesses and interference terms in the intermediate orders.
The dominant resonant $t\bar t$ production channel appears at $\mathcal{O}(\alpha_s^2\alpha^4)$, while the highest-power QCD term $\mathcal{O}(\alpha_s^4\alpha^2)$ acts as an irreducible QCD background with no resonant top quarks and at most one resonant $W$ boson. At NLO, the complete correction comprises six perturbative orders from $\mathcal{O}(\alpha_s^5\alpha^2)$ down to $\mathcal{O}(\alpha^7)$, including QCD, EW, and mixed QCD-EW contributions. A crucial point is that in the complete calculation, QCD and EW effects cannot be cleanly separated into two independent NLO QCD and NLO EW computations, since mixed interferences are required for the consistent cancellation of soft and collinear singularities across the full set of real and virtual contributions.
Tree-level and one-loop amplitudes are generated and evaluated with \textsc{Recola} \cite{Actis:2012qn}. One-loop tensor integrals are computed with \textsc{Collier} \cite{Denner:2016kdg}. Phase-space integration is performed with adaptive multi-channel techniques using \textsc{Parni} \cite{vanHameren:2007pt} and \textsc{Kaleu} \cite{vanHameren:2010gg}. Real-emission contributions and subtraction of soft/collinear singularities are handled within \textsc{Helac-Dipoles} \cite{Czakon:2009ss} using both Nagy-Soper \cite{Bevilacqua:2013iha} and Catani-Seymour \cite{Catani:1996vz} formalisms, extended to photon radiation following Ref.~\cite{Dittmaier:1999mb}.
Including NLO EW effects requires a jet definition that remains infrared safe in the presence of photons. We employ democratic photon-parton clustering, where photons and QCD partons are clustered on equal footing. A clustered object is classified as an isolated photon (photon-jet) if the photon energy fraction exceeds a cut $z_{\gamma,\text{cut}}$, and as a hadronic jet otherwise, with
\begin{equation}
z_\gamma=\frac{p_{T,\gamma}}{p_{T,\gamma}+p_{T,i}}\,, \qquad i=q,\bar q,g\,.
\end{equation}
Imposing such a classification renders the prediction sensitive to the final-state collinear region of $q\to q\gamma$ splittings. To restore IR safety, the remaining collinear singularities are absorbed into a parton-to-photon fragmentation function \cite{Glover:1993xc}, implemented consistently in our subtraction framework.

In addition, mixed QCD-EW corrections contain configurations where a photon splits collinearly, $\gamma\to q\bar q$, and the resulting pair is identified as a light jet by the jet algorithm. The corresponding singular contribution would cancel against EW virtual corrections to an underlying configuration with an external photon, but that configuration is excluded by the definition of our process, which requires two light jets. We therefore factorize the leftover collinear divergence into a photon-to-jet conversion function $D_{\gamma\to\text{jet}}$ defined in Ref.~\cite{Denner:2019zfp}, which acts as a non-perturbative counterterm and renders the complete prediction finite in the presence of $\gamma\to q\bar q$ splittings.
A resonance-enhancing cut $|M_{jj}-m_W|<{\cal Q}_{\text{cut}}$ (${\cal Q}_{\text{cut}}=20~\mathrm{GeV}$) is applied to select light-jet pairs compatible with a hadronic $W$ decay.

\section{Integrated fiducial cross sections}
\label{sec:integrated}\begin{table}[t!]
    \centering
    \renewcommand{\arraystretch}{1.2}
    \begin{tabular}{ll@{\hskip 10mm}l@{\hskip 10mm}l@{\hskip 10mm}}
        \hline
        \noalign{\smallskip}
         &&$\sigma_{i}$ [pb] & Ratio to ${\rm Born}_3$  \\
        \noalign{\smallskip}\midrule[0.5mm]\noalign{\smallskip}
        \Bornone&\asLOone& $ 0.138(2) $ & $ 0.44\% $ \\
        \Borntwo&\asLOtwo& $ 0.0002(1) $ & $ 0.00\% $ \\
        \Bornthree&\asLOthree& $ 31.264(4) $ & $ 100.00\% $ \\
        \Bornfour&\asLOfour& $ 0.1336(1) $ & $ 0.43\% $ \\
        \Bornfive&\asLOfive& $ 0.0968(1) $ & $ 0.31\% $ \\
        \noalign{\smallskip}\hline\noalign{\smallskip}
        \NLOone&\asNLOone& $ +\,0.144(4) $ & $ +\,0.46\% $\\
        \NLOtwo&\asNLOtwo& $ -\,0.0022(4) $ & $ -\,0.01\% $\\
        \NLOthree&\asNLOthree& $ -\,1.90(2) $ & $ -\,6.10\% $\\
        \NLOfour&\asNLOfour& $ +\,0.112(2) $ & $ +\,0.36\% $\\
        \NLOfive&\asNLOfive& $ +\,0.1864(6) $ & $ +\,0.60\% $\\
        \NLOsix&\asNLOsix& $ +\,0.00228(2) $ & $ +\,0.01\% $\\
        \noalign{\smallskip}\hline\noalign{\smallskip}
        \LOfull & $\sum_{i=1}^5 \text{LO}_i$ & $ 27.241(2)^{+30.7\%}_{-22.0\%} $ & $ 0.87 $ \\
        \Bornfull& $\sum_{i=1}^5 \text{Born}_i$ & $ 31.634(4)^{+30.7\%}_{-22.0\%} $ & $ 1.01 $ \\
        \NLOqcd& $\text{Born}+\text{NLO}_3$ & $ 29.72(2)^{+1.4\%}_{-6.0\%} $ & $ 0.95 $ \\
        \NLOfull& $\text{Born}+\sum_{i=1}^6 \text{NLO}_i$ & $ 30.16(2)^{+1.5\%}_{-5.6\%} $ & $ 0.96 $ \\
        \noalign{\smallskip}\hline\noalign{\smallskip}
    \end{tabular}
    \caption{Integrated fiducial cross section with the resonance-enhancing requirement
$|M_{jj}-m_W|<{\cal Q}_{\text{cut}}$ with ${\cal Q}_{\text{cut}}=20$ GeV.
Numbers in parentheses indicate Monte Carlo integration uncertainties.}
    \label{tab:integratedqcut}
\end{table}

Table~\ref{tab:integratedqcut} shows the perturbative decomposition of the integrated fiducial cross section. The hierarchy is dominated by $\text{Born}_3$, while $\text{Born}_1$ provides the leading irreducible QCD background component which is at the percent level compared to $\text{Born}_3$. At NLO, $\text{NLO}_3$ controls the size of the higher-order correction, whereas all other $\text{NLO}_i$ terms are individually at the sub-percent level relative to $\text{Born}_3$. Nevertheless, they are required for the completeness and IR finiteness of the combined QCD and EW prediction. We note that without imposing the $M_{jj}$ window, the integrated NLO correction is substantially larger and the background-like $\text{Born}_1$ contribution is enhanced.

\section{Differential distributions}

\begin{figure}[t!]
    \begin{center}
    \includegraphics[width=0.49\textwidth]{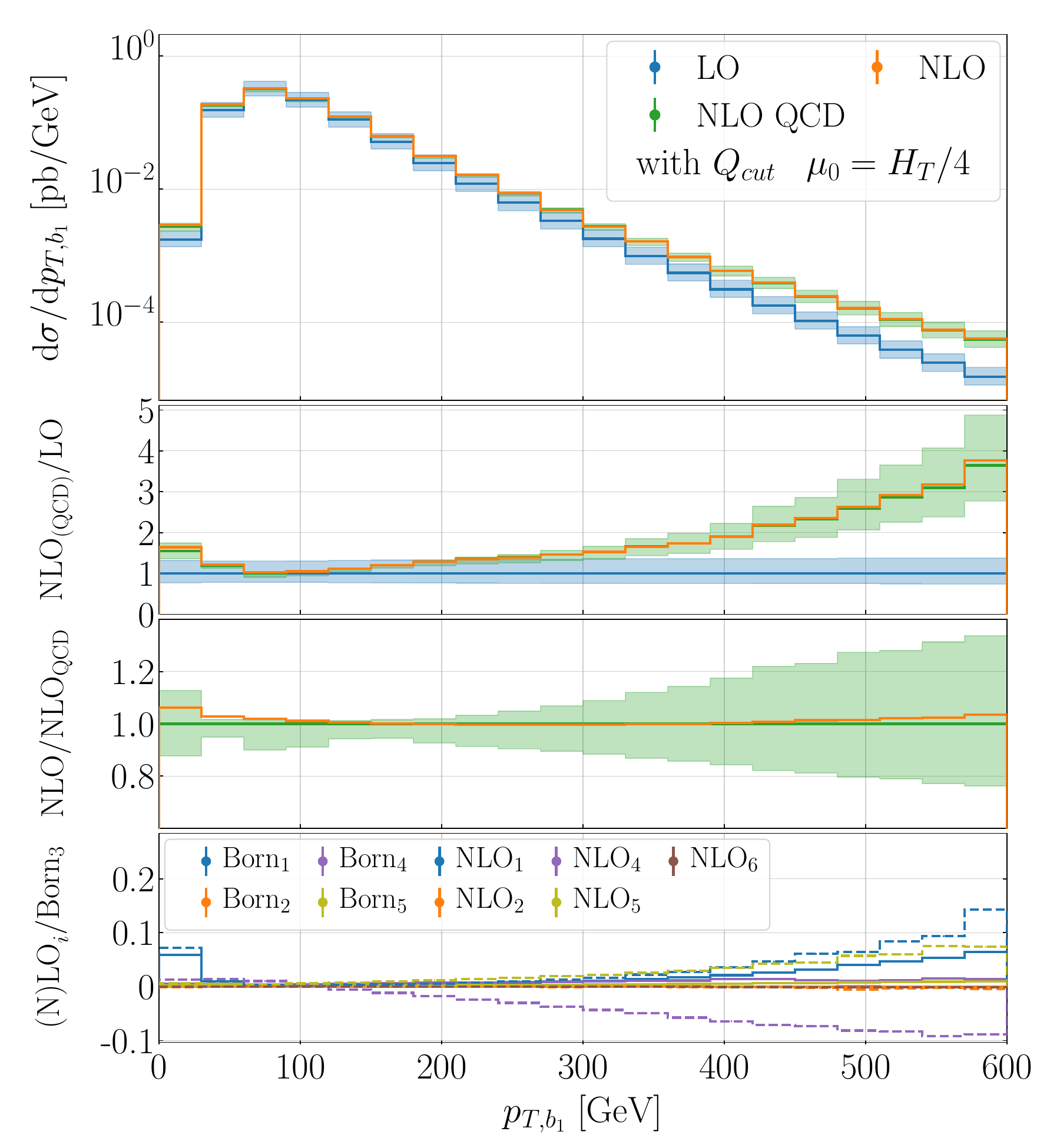}
    \includegraphics[width=0.49\textwidth]{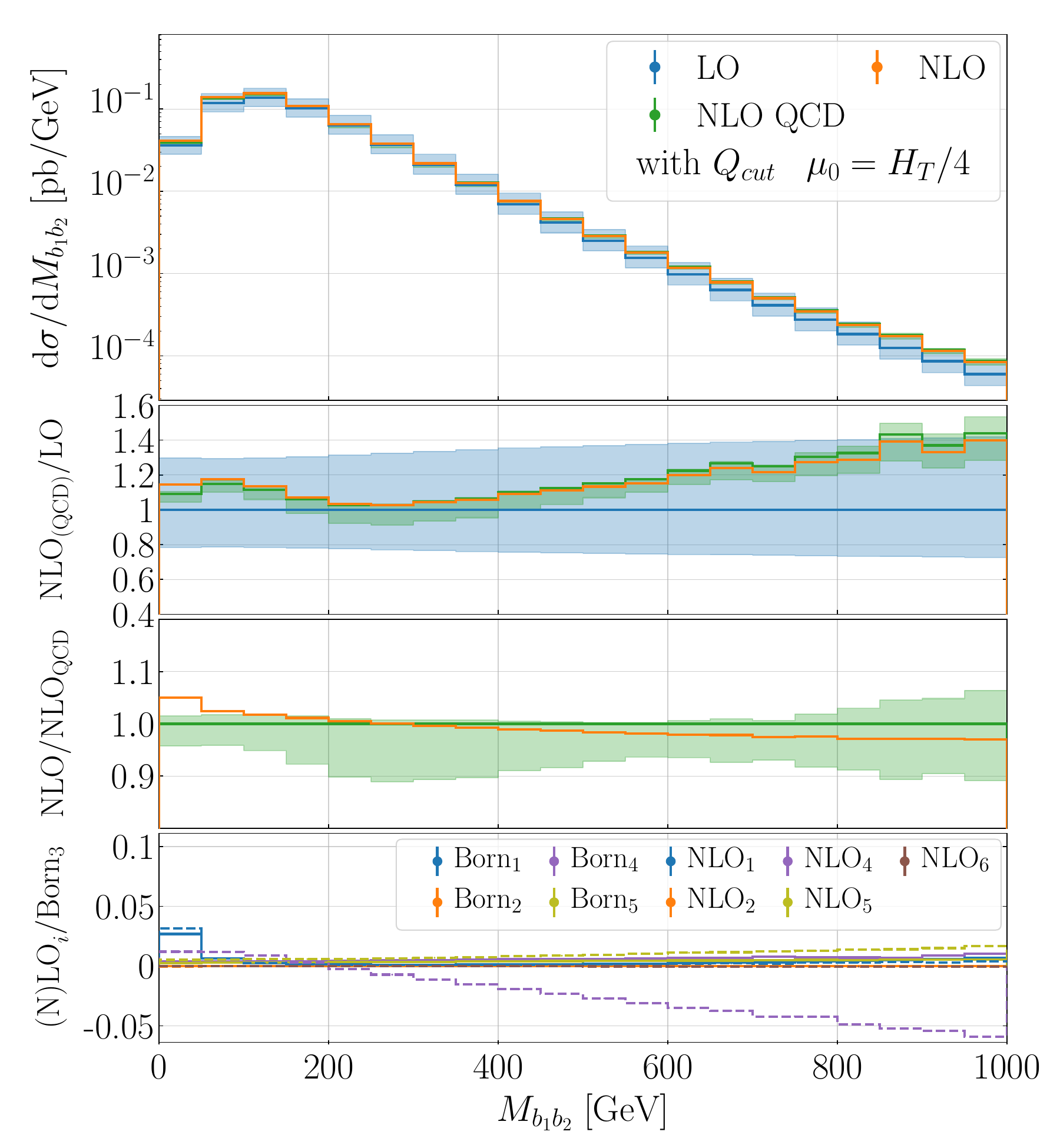}
    \end{center}
    \caption{\it Differential distributions for $p_{T,b_1}$ (left) and $M_{b_1b_2}$ (right) at LO, $\mathrm{NLO}_{\mathrm{QCD}}$, and complete NLO, shown with ratio panels.}
    \label{fig:diff}
\end{figure}

We focus on two representative observables: the transverse momentum of the hardest $b$-jet, $p_{T,b_1}$, and the invariant mass of the $b$-jet pair, $M_{b_1b_2}$, shown in Figure \ref{fig:diff}. They are dimensionful observables that are sensitive to additional QCD radiation, non-resonant backgrounds, and EW Sudakov effects. For $p_{T,b_1}$, the high-$p_T$ tail receives very large positive NLO corrections. In this region, non-resonant components become visible, with $\text{Born}_1$ and its QCD correction $\text{NLO}_1$ reaching the $\mathcal{O}(10\%)$ level relative to $\text{Born}_3$. EW Sudakov effects in $\text{NLO}_4$ provide a negative contribution of comparable size, partially cancelling positive subleading terms such as $\text{NLO}_5$. For $M_{b_1b_2}$, the NLO enhancement is more moderate than for $p_{T,b_1}$. The dominant subleading effect is the negative $\text{NLO}_4$ contribution, at the $\mathcal{O}(10\%)$ level in the high-mass tail, which is caused by EW Sudakov suppression, while other subleading orders remain small.

\section{Conclusions}
We presented complete NLO predictions for full off-shell $t\bar t$ production in the $\ell +j$ channel, including all Born-level orders, all NLO perturbative contributions, and all resonance topologies with interferences. A key technical component is IR safety in the presence of photons and jets, achieved via democratic clustering with a photon-energy fraction criterion, a parton-to-photon fragmentation function, and a photon-to-jet conversion function accounting for $\gamma\to q\bar q$ contributions. Phenomenologically, while subleading NLO terms are small at the integrated fiducial cross-section level, they can reach the $\mathcal{O}(10\%)$ level in the high-$p_T$ tails and can partially cancel. This motivates the use of complete NLO predictions for precision applications at the LHC.

\section*{Acknowledgments}
This work was supported by the DFG under grant 396021762 - TRR 257: P3H - \textit{Particle Physics Phenomenology after the Higgs Discovery} and grant 400140256 - GRK 2497: \textit{The Physics of the Heaviest Particles at the LHC.}

\bibliographystyle{elsarticle-num} 

\begin{thebibliography}{10}
\expandafter\ifx\csname url\endcsname\relax
  \def\url#1{\texttt{#1}}\fi
\expandafter\ifx\csname urlprefix\endcsname\relax\def\urlprefix{URL }\fi
\expandafter\ifx\csname href\endcsname\relax
  \def\href#1#2{#2} \def\path#1{#1}\fi

\bibitem{Denner:2017kzu}
A.~Denner, M.~Pellen, \textit{Off-shell production of top-antitop pairs in the  $\ell+jets$ channel at NLO QCD}, JHEP 02 (2018) 013.

\bibitem{Mans:2025czh}
L.~Mans, D.~Stremmer, M.~Worek, \textit{Complete NLO corrections to full off-shell
  $t\bar{t}$ production in the $\ell + j$ decay channel}, \texttt{arXiv:2512.18391}.

\bibitem{Denner:2014bna}
A.~Denner, S.~Dittmaier, M.~Hecht, C.~Pasold, \textit{NLO QCD and electroweak
  corrections to $W+\gamma$ production with leptonic W-boson decays}, JHEP 04
  (2015) 018.

\bibitem{Stremmer:2024zhd}
D.~Stremmer, M.~Worek, \textit{NLO QCD predictions for $ t\overline{t}\gamma $ with
  realistic photon isolation}, JHEP 01 (2025) 156.

\bibitem{Denner:2019zfp}
A.~Denner, S.~Dittmaier, M.~Pellen, C.~Schwan, \textit{Low-virtuality photon
  transitions $\gamma^*\to f\bar f$ and the photon-to-jet conversion function},
  Phys. Lett. B 798 (2019) 134951.

\bibitem{Actis:2012qn}
S.~Actis, A.~Denner, L.~Hofer, A.~Scharf, S.~Uccirati, \textit{Recursive generation of
  one-loop amplitudes in the Standard Model}, JHEP 04 (2013) 037.

\bibitem{Denner:2016kdg}
A.~Denner, S.~Dittmaier, L.~Hofer, \textit{Collier: a fortran-based Complex One-Loop
  LIbrary in Extended Regularizations}, Comput. Phys. Commun. 212 (2017)
  220--238.

\bibitem{vanHameren:2007pt}
A.~van Hameren, \textit{PARNI for importance sampling and density estimation}, Acta
  Phys. Polon. B 40 (2009) 259--272.

\bibitem{vanHameren:2010gg}
A.~van Hameren, \textit{Kaleu: A General-Purpose Parton-Level Phase Space Generator}
  (3 2010).

\bibitem{Czakon:2009ss}
M.~Czakon, C.~G. Papadopoulos, M.~Worek, \textit{Polarizing the Dipoles}, JHEP 08
  (2009) 085.

\bibitem{Bevilacqua:2013iha}
G.~Bevilacqua, M.~Czakon, M.~Kubocz, M.~Worek, \textit{Complete Nagy-Soper subtraction
  for next-to-leading order calculations in QCD}, JHEP 10 (2013) 204.

\bibitem{Catani:1996vz}
S.~Catani, M.~H. Seymour, \textit{A General algorithm for calculating jet
  cross-sections in NLO QCD}, Nucl. Phys. B 485 (1997) 291--419, [Erratum:
  Nucl.Phys.B 510, 503--504 (1998)].

\bibitem{Dittmaier:1999mb}
S.~Dittmaier, \textit{A General approach to photon radiation off fermions}, Nucl.
  Phys. B 565 (2000) 69--122.

\bibitem{Glover:1993xc}
E.~W.~N. Glover, A.~G. Morgan, \textit{Measuring the photon fragmentation function at
  LEP}, Z. Phys. C 62 (1994) 311--322.

\end{thebibliography}

\end{document}